\documentclass[12pt,letterpaper,onecolumn]{article}
\usepackage[latin1]{inputenc}
\usepackage{amsmath}
\usepackage{amsfonts}
\usepackage{amssymb}
\usepackage{braket}
\usepackage{enumerate}
\usepackage{graphicx}

\def\sb{{\sf b}}

\def\cH{{\cal H}}

\def\tr{{\rm tr}}

\newcommand\ketbra[2]{|#1\rangle\langle #2|}

\newcommand\trnorm[1]{\| #1 \|_1}

\usepackage[left=1.00in, right=1.00in, top=1.00in, bottom=1.00in]{geometry}

\author{\\
\\
Horace P. Yuen\\
Department of Electrical Engineering and Computer Science\\
Department of Physics and Astronomy\\
Northwestern University, Evanston Il. 60208\\
yuen@eecs.northwestern.edu
}
\title{An Unconditionally Secure Quantum Bit Commitment Protocol}

\begin{document}
\linespread{1}
\maketitle
\linespread{2}

\begin{abstract}
This article describes a quantum bit commitment protocol, QBC1, based on entanglement destruction via forced measurements and proves its unconditional
security.
\end{abstract}

\vspace{10mm}

\noindent\textbf{Note:} This paper is an elaboration of my 2006 QCMC paper, arXiv: 0702074v4(2007), published in its Proceedings volume. It was submitted in Dec 2009 as an ``invited paper'' to a journal, which was withdrawn half a year later because the editors found it incomprehensible. I hope it may make better sense to some other readers.

My reasons for the impossibility of QBC ``impossibility proofs'' are described in ref [1]. Over the years I have produced several QBC protocols that I thought were secure, but when concealing they are not binding due to the scope of entanglement attack that works even across teleportation. I did not and do not see such scope spelled out anywhere, though before putting such papers on the arXiv I should have tried harder to find out whether entanglement attack works in my cases, which I eventually did. Since 2003 I have not received any substantial negative comment on my QBC arXiv papers, only getting a few questions and agreements, and thus the arXiv papers have not served the purpose of soliciting technical disagreements I sought in this controversial subject.

I have been as sure that the present protocol is secure as most results I ever published, but I knew the environment of disagreement and did not submit any QBC paper to any journal until Dec 2009. If this present paper is indeed incomprehensible, it would have to be expanded before submission to a journal. In the meantime a QBC possibility paper by G. P. He, J. Phys. A: Math. Theor. 44, 445305 (2011) has appeared in a reputable journal. That protocol is based on an entirely different mechanism from that of this paper, and gives a weaker form of security. Generally, the best a QBC impossibility proof
can do is to show a certain type of QBC protocols cannot be unconditionally secure. It cannot show general impossibility for the simple reason that not all QBC protocols can be captured in any mathematical formulation just within nonrelativistic quantum mechanics [1].

My view is that QBC can actually be practically developed and it could perform cryptographic functions with security that is impossible to achieve classically. However, it would not be through the impractical protocol of this paper and the security would not be ``unconditional'' which is never needed in practice. It appears such QBC development is only possible after the entrenched contrary view on unconditionally secure QBC is sufficiently softened up. I hope this paper would contribute to such end.

\clearpage

\section{Introduction}\label{sec:intro}

It is nearly universally accepted that unconditionally secure quantum
bit commitment (QBC) is impossible. This is taken to be a consequence of
the Einstein-Podolsky-Rosen(EPR) type entanglment cheating. For detailed
discussion with historical remarks on the impossibility of secure QBC and
the various impossibility proofs, see ref \cite{yuen1}-\cite{dariano07}.
In the following, a new approach is described that lies outside the formulation of
these impossible proofs. A secure QBC protocol, to be called QBC1, is presented
together with a full proof of its unconditional security. This paper is completely self-contained other than background knowledge of quantum mechanics.

\section{QBC Formulation and the Impossibility Proof}\label{sec:form}

In a {\it bit commitment} scheme, one party, Alice,
provides another party, Bob, with a piece of evidence that she has
chosen a bit \sb\ (0 or 1) which is committed to him.  Later, Alice
would {\it open} the commitment by revealing the bit \sb\ to Bob and
convincing him that it is indeed the committed bit with the evidence
in his possession and whatever further evidence Alice then provides,
which he can {\it verify}.  The usual concrete example is for Alice to
write down the bit on a piece of paper, which is then locked in a safe
to be given to Bob, while keeping for herself the safe key that can
be presented later to open the commitment.  The scheme should be {\it
binding}, i.e., after Bob receives his evidence corresponding to a given bit
value, Alice should not be able to open a different one and convince
Bob to accept it. It should also be {\it concealing}, i.e., Bob
should not be able to tell from his evidence what the bit \sb\ is.
Otherwise, either Alice or Bob would be able to cheat successfully.

In standard cryptography, secure bit commitment is to be achieved
either through a trusted third party, or by invoking an unproved
assumption concerning the complexity of certain computational
problems.  By utilizing quantum effects, specifically the intrinsic
uncertainty of a quantum state, various QBC schemes not
involving a third party have been proposed to be unconditionally
secure, in the sense that neither Alice nor Bob could cheat with any
significant probability of success as a matter of physical laws.  In
1995-1996, a supposedly general proof on the impossibility of
unconditionally secure QBC and the insecurity of previously proposed
protocols were presented \cite{may1}-\cite{lc1}.  Henceforth it has
been generally accepted that secure QBC and related objectives are
impossible as a matter of principle \cite{lc2}-\cite{sr}.

There is basically just one impossibility proof, which gives the EPR
attacks for the cases of equal and nearly equal density operators that Bob
has for the two different bit values.  The proof purports to show that
if Bob's successful cheating probability $P^B_c$ is close to the
value $\frac{1}{2}$, which is obtainable from pure guessing of the bit value,
then Alice's successful cheating probability $P^A_c$ is close to the
perfect value 1. The impossibility proof
describes the EPR attack on a specific type of protocols, and then
argues that all possible QBC protocols are of this type.

The formulation of the standard impossibility proof can be cast as
follows.  Alice and Bob have available to them two-way quantum
communications that terminate in a finite number of exchanges, during
which either party can perform any operation allowed by the laws of
quantum physics, all processes ideally accomplished with no
imperfection of any kind.  During these exchanges, Alice would have
committed a bit with associated evidence to Bob.  It is argued that,
at the end of the commitment phase, there is an openly known entangled pure state
$\ket{\Phi_\sb}$, $\sb \in \{0,1\}$, shared between Alice who
possesses state space $\cH^A$, and Bob who possesses $\cH^B$.  For
example, if Alice sends Bob one of $M$ possible states $\{
\ket{\phi_{\sb i}} \}$ for bit \sb\ with probability $p_{\sb i}$, then
\begin{equation}
\ket{\Phi_{\sb }} = \sum_i \sqrt{p_{\sb i}}\ket{e_i}\ket{\phi_{\sb i}}
\label{eq:1}
\end{equation}
with orthonormal $\ket{e_i} \in \cH^A$ and known $\ket{\phi_{\sb i}}
\in \cH^B$.  Alice would open by making a measurement on $\cH^A$, say
$\{ \ket{e_i} \}$, communicating to Bob her result $i_0$, then Bob would verify by measuring
the corresponding projector $\ketbra{\phi_{\sb i_0}}{\phi_{\sb i_0}}$ on $\cH^B$.

When classical random numbers known only to one party
are used in the commitment, they are to be replaced by corresponding
quantum entanglement purification.  The commitment of $\ket{\phi_{\sb i}}$
with probability $p_{\sb i}$ in (\ref{eq:1}) is, in
fact, an example of such purification. Generally, for any random $k$ used by Bob, it is argued from the
doctrine of the ``Church of the Larger Hilbert Space'' that it is to be
replaced by the purification $\ket{\Psi}$ in $\cH^{B_1} \otimes \cH^{B_2}$,
\begin{equation}
\ket{\Psi} = \sum_k \sqrt{\lambda_k} \ket{f_k}\ket{\psi_k},
\label{eq:2}
\end{equation}
where $\ket{\psi_k} \in \cH^{B_2}$. The \{$\ket{f_k}$\} are complete orthonormal in $\cH^{B_1}$ kept
by Bob while $\cH^{B_2}$ would be sent to Alice.

For unconditional, rather than perfect, security, one demands that
both cheating probabilities $P^B_c - \frac{1}{2}$ and $P^A_c$ can
be made arbitarily small when a security parameter $n$ is increased
\cite{may1}.  Thus, {\it unconditional security} is quantitatively expressed
as
\begin{equation}
\qquad \lim_n P^B_c = \frac{1}{2},\quad \lim_n P^A_c = 0.
\label{eq:3}
\end{equation}
The condition (\ref{eq:3}) says that, for any $\epsilon > 0$, there
exists an $n_0$ such that for all $n > n_0$, $P^B_c - \frac{1}{2} <
\epsilon$ and $P^A_c < \epsilon$, to which we may refer as
$\epsilon$-{\it concealing} and $\epsilon$-{\it binding}.  These
cheating probabilities are to be computed purely on the basis of
logical and physical laws, and thus would survive any change in
technology, including an increase in computational power.  In general,
one can write down explicitly the optimal $P^B_c$,
\begin{equation}
P^B_c = \frac{1}{4}\left(2 + \trnorm{\rho^B_0 - \rho^B_1}\right),
\label{eq:4}
\end{equation}
where $\trnorm{\cdot}$ is the trace norm, $\trnorm{\tau} \equiv \tr
(\tau^\dag \tau)^{1/2}$ for a trace-class operator $\tau$.

The entanglement cheating mechanism is explicitly spelled out in the impossibility proof.
Under perfect concealing $P^B_c=\frac{1}{2}$, it follows from (\ref{eq:4}) that the state $\rho^B_b$ at Bob's possession
obeys $\rho^B_0=\rho^B_1$. Hence by the Schmidt decomposition Alice can turn $\ket{\phi_{0i}}$ into $\ket{\phi_{1i}}$ by a unitary
transformation on $\cH^A$ in her possession, thus succeeds in cheating perfectly. Under approximate concealing,
an explicit transformation on $\cH^A$ can be similarly identified \cite{yuen2}-\cite{yuec} which leads to
\begin{equation}
4(1-P^B_c)^2 \le P^A_c \le 2 \sqrt{P^B_c
  (1-P^B_c)}.
\label{eq:5}
\end{equation}
The lower bound in (\ref{eq:5}) yields the following
impossibility result,
\begin{equation}
\lim_n P^B_c = \frac{1}{2} \,\, \Rightarrow
  \,\, \lim_n P^A_c = 1
\label{eq:6}
\end{equation}
Note that the impossibility
proof makes a stronger statement than the mere
impossibility of unconditional security, i.e., (\ref{eq:6}) is stronger than
(\ref{eq:3}) not being possible.

The assumption in the impossibility formulation that $\ket{\Phi_{\sb }}$ are openly known
has been challenged. In a multi-pass protocol where Alice and Bob exchange states, each $\ket{\phi_{\sb i}}$
becomes of the form $\ket{\phi_{\sb i k}}$ \cite{sr}
\begin{equation}
\ket{\phi_{\sb ik}} = U^A_{\sb i_n} \ldots U^A_{\sb i_2} U^B_{k_1}
U^A_{\sb i_1} \ket{\phi_0}.
\label{eq:7}
\end{equation}
where $U^A_{\sb il}$ are unitaries that Alice applies and $U^B_{\sb kj}$ are applied by Bob. The ancilla
state $\ket{e_i}$ also separates into $\ket{e^A_i}\ket{e^B_k}$ with $\ket{e^A_i}$ in Alice's possession and
$\ket{e^B_k}$ in Bob's. It is clear that the exact $\ket{e^B_k}$ may be kept secret by Bob, in an unnormalized
form that would include both the entanglement basis and the probability of each state in it. The question is why secure
QBC is impossible under such added randomness, whose quantum purification is either unknown to anyone as in the case of classical random
number generation from a piece of macroscopic equipment, or at least known only to the party who preforms the entanglement
purification.

For some discussion of this point of employing unknown randomness, see \cite{yuec}-\cite{yueb2} and references cited
therein. It turnes out it appears impossible to get a secure protocol with this approach. For the case of perfect concealing, a
general proof of this impossibility was given in \cite{yuen2} for a two-pass protocol. A different argument applicable
to multi-pass protocol was given by Ozawa\cite{oz} and later independently by Cheung\cite{chau1}. Simple as well as more complicated
proofs concerning all natural protocols of this kind in the case of approximate concealing are also available. See \cite{yuen1}-\cite{dariano07},
 \cite{chau2}.

In the above formulation one may consider, \textit{more generally}, the whole $\ket{\Phi_{\sb }}$ of (1) as the state corresponding to the bit \sb\
with Alice sending $\cH^A$ to Bob at opening who verifies by measuring on the total $\ket{\Phi_{\sb }}$.
Similarly in the multi-pass case, (7) is generalized from $\ket{\phi_{\sb ik}}$ to $\ket{\Phi_{\sb ik}}$
with different subspaces of $\cH^A$ and $\cH^B$ being exchanged during each pass.
The above quantitative conclusion is not affected. Note, however, that either Alice or Bob has to provide the
initial state $\ket{\phi_0}$. Indeed, $\ket{\phi_0}$ must be on a large enough dimension state space and openly known to both parties if either can
perform random number purification. It is more convenient to just let each party supply its own state space at each turn when needed,
and let $\cH^A$ and $\cH^B$ be their individual total spaces as just indicated. In contrast to one-pass protocols in \cite{lc1}, there is then always the question
of ``\textit{honesty}'' in multi-pass protocols. It is clear that some form of state checking may be necessary to execute these protocols.

\section{New Approach}\label{sec:newappr}
In the impossibility proof formulation the probability of interactive checking between Alice and Bob, similar to
there in QKD protocol such as BB84, is not explicitly accounted for. Even if Bob's check on Alice can be postponed to just before
opening, Alice's check on Bob must be carried out during the commitment phase to maintain $\epsilon$-concealing, The implicit assumption
must be, therefore, that such checking could be satisfied perfectly without affecting the protocol. In this section, a new approach to QBC
protocol would be described that shows such implicit assumption cannot be true. This approach would be utilized in the next Section \ref{sec:qbc1}
to show how a specific secure protocol can be obtained.

Consider the following situation or ``protocol": Bob sends Alice a sequence of $n$ qubits, each randomly in one
of the two orthogonal states $\ket{l_j}, j \in \{1,2\}$, which are themselves chosen randomly on a fixed great circle $C$ of
the qubit Bloch sphere. The index $l$ indicates the position in the $n$-qubit sequence. We assume for convenience that
Bob entangled each $l$th qubit to a qubit ancilla he keeps. Alice randomly picks one $\ket{\bar{l}}$,
modulates it by  $U_{0}=R(\frac{\pi}{2})$ or $U_{1}=R(-\frac{\pi}{2})$, rotation
by two different angles on $C$, depending on $\sb\ \in \{0,1\}$, and sends it back to Bob as
commitment. Alice opens by sending back the rest and revealing
everything. Let $|k\rangle \in
\cH^{A}$ be the orthogonal entanglement ancilla states, $P$ the cyclic shift
unitary operator on $n$ qubits, $P^n=I$. Suppose Alice
entangles in a minimal way,
\begin{equation}\label{eq:8}
|\Psi_{\sb}\rangle=U_{\sb}\frac{1}{\sqrt{n}}\sum_{k=1}^n|k\rangle\otimes P^{k}|1_{j}\rangle ...|n_{j}\rangle
\end{equation}
where $|1_{j}\rangle$ is acted on by $U_{\sb}$.  This ``protocol'' can be
shown to be $\epsilon$-concealing, and Alice can locally turn $|\Psi_{0}\rangle$ to
$|\Psi_{1}\rangle$ near perfectly in a standard entanglement cheating.

Consider a protocol with the following added checking to the above. Before opening, Bob asks Alice to
send back a fraction $\lambda$, say $\lambda=\frac{1}{2}$, of the $n$ qubits chosen randomly by Bob for checking. If Alice replies that
fraction contains the committed one, Bob would ask to check the remaining $1-\lambda$ fraction instead. Assuming
Alice has to answer correctly, she must measure on $\cH^A$ to get a specific $\ket{k}$. After Bob's checking, he still
has a uniform distribution on exactly what the original committed qubit is according to his own positions. Thus the protocol \textit{remains}
$\epsilon$-concealing if $n$ is sufficiently large, while Alice has lost her entanglement cheating capability. This is what
was referred to as ``the destruction of entanglement for cheating" in several of my previous protocols, beginning with a first one at
the 2000 QCMC meeting in Capri, Italy.

Such ploy did not lead to a secure protocol because the entanglement (\ref{eq:8}) or a similar sparsely entangled one was not insisted upon as part of
the protocol prescription. Before it will be discussed in the following how the entanglement (\ref{eq:8}) can be enforced, note that Alice can retain her
entanglement cheating capability by other entanglements, in particular by the full $n$-permutation group. She could name her entanglement
basis vectors $\ket{k}$ by the original positions of the qubits Bob sent,
\begin{equation}
\label{eq:9}
\ket{k}\rightarrow\ket{1(k_1),\ldots,n(k_n)}
\end{equation}
where $l(k_l)$ indicated that original qubit $l$ is at position $k_l$ corresponding to $\ket{k}$. When she is asked to return a fraction
$\lambda$ that has positions $\lambda(m)$, $m\in\overline{1-n}$,  she would perform a L\"{u}ders measurement, that is, a projection $P^{\prime}$ into
the subspace in $\cH^A$ that fixes the \{$\lambda(m)$\} position. If the entanglement is sufficient dense, the remaining $1-\lambda$ fraction is still entangled
in the remaining ancilla space $(1-P^{\prime})\cH^A$, and entanglement cheating remains possible. With the entanglement (\ref{eq:8}) there is no
such degeneracy. In fact, $P^{\prime}\cH^A=\cH^A$. Thus, fixing the position of just one qubit already fixes the positions of all the others.

Note that the checking of ancilla is naturally included in the generalized formulation discussed in the last paragraph of section
\ref{sec:form}---there is no system or subsystem that cannot be exchanged. The question now is why Alice should entangle as in (\ref{eq:8})
rather than one which allows her to cheat later. In the QBC literature, with the possible excepting of (\ref{eq:2}), the claim has always
been that even under \textit{honest} following of the protocol prescription, no protocol can be secure \cite{lc2}-\cite{sr}.
In the presence of interactive checking as above, we here conclusively shown that such claim is incorrect.

The ``honesty" assumption is widely used in the literature to describe multi-pass protocol including those for quantum coin tossing \cite{dk}.
It may or may not make sense depending on whether the``honest" action can in principle be checked by the other party without rendering the
protocol ineffective. For example, in the simple one-pass protocol of \cite{lc1}, it makes no sense to require Alice to be honest and does \textit{not}
entangle. It is clear that an actual physical entanglement is needed for the EPR cheating even when the protocol is perfectly concealing. Note that this is in fact
the \textit{basis} of the success of checking for preventing entanglement cheating with (\ref{eq:8}), that only classical randomness is left
after checking. Thus, Alice would entangle
anyway in the situation of \cite{lc1} and the simple protocol that requires such ``honesty'' is not secure.

In a multi-pass protocol, there is always the question whether ``honest entanglement'' or any other prescription of the protocol is followed. Even with just a
two-pass QBC protocol in which Bob first sends Alice some qubits in prescribed states, including the above ``protocol'' involving (\ref{eq:8}), he
can easily cheat by sending in other qubits instead. For example, he could send in identical fixed qubit states and so he would know how to measure
to distinguish $\sb=0,1$ from the committed qubit with considerable $P^B_c>\frac{1}{2}$ for any $\{U_0, U_1\}$ pair.
He is prevented from such cheating via checking of one form or another.

A crucial question is: what happens when one party is found cheating during protocol execution. Clearly the party cannot be allowed to keep cheating
indefinitely, if only because of ``intent'' \cite{yuec} since the party does not need to participate to begin with. I have previously described \cite{app} several
approaches to deal with this problem which has \textit{not} yet received an adequate discussion in the literature, but which can be solved in one stroke by
an honesty assumption that requires all the parties to be perfectly honest in their prescribed actions and thus no cheating would even be found before opening.
This is a perfectly reasonable working assumption for the ideal protocol under discussion \textit{as long as} the action can be checked, in view of the discussion
just given above. It is equivalent to the assignment of infinite penalty in a game type formulation \cite{app}, and it allows us to bring forth our
new point without the burden of technicalities. It is also exactly what has been implicitly assumed in the literature as we mentioned.

Note that the whole protocol may need to be started all over again after a checking. It is easy to see that in the absence of
resource constraint as in the case of all QBC impossibility proof formulations thus far, one party
can check the same state an arbitrarily large number of times before proceeding. The total number of checks may grow multiplicatively, not just linearly, with the
number of state checking.
It is reasonable to count cheating detection probability as the party's failure probability in $P^A_c$ and
$P^B_c$. Thus, whenever a bound is imposed on the allowable total number of cheatings getting caught, an unconditionally secure protocol would be
obtained which is equivalent to the honest assumption. This is because both $P^A_c$ and $P^B_c$ can be brought arbitrarily close to
their prescribed $\epsilon$-level with a large enough number of checkings on each state.

The point that was made in this section in connection with (\ref{eq:8}) has the following \textit{general implication} independently
of whether a secure protocol can be made on that basis:  There is no general impossibility proof that shows the entanglement
formed by one party as prescribed by a QBC protocl would have effective remaining entanglement after
checking. In the next section, however, we do exhibit such a specific secure protocol.

\section{Secure Protocol QBC1}\label{sec:qbc1}
We consider the following protocol QBC1 \cite{qbc3} in which Bob sends Alice a sequence of $n$ qubits as described in the last section,
requiring Alice to entangle as in (\ref{eq:8}). We will show later in appropriate places how that as well as any other prescribed states
for Alice and Bob can be checked. That the protocol is $\epsilon$-concealing is intuitively obvious, and can be proved as follow.
For simplicity we let the protocol prescribe that each of the qubit state Bob sends is entangled with an ancilla in his possession.
Alice can check this before proceeding by asking Bob to send her the qubit ancilla and measuring to verify.
\\

\noindent Concealing Proof for QBC1:

First we assume that Bob does not permutation entangle the $n$ qubit. It is technically messy to show concealing if she does,
but the absence of such permutation entanglement can be assured by requiring Bob to permutation entangle as in (\ref{eq:8}),
and destroyed by Alice asking to check one or more of the qubits. That Bob did entangle in such manner in the first place can be checked by asking
him to send in the ancilla for Alice to check.

For simplicity we do not distinguish here a qubit state from the qubit which is clear from context. Let $a_l$ be the ancilla part
of Bob's states entangled to the $l$th qubit. Then $a_l=\frac{I}{2}$ without the $i$th qubit and
\begin{equation}
\label{eq:10}
\rho_\sb=\frac{1}{n}\sum_{l=1}^n a_1\otimes\ldots\otimes(\sigma_\sb a_l)\otimes\cdots\otimes a_n
\end{equation}

\noindent In (10), $(\sigma_\sb a_ l)$ denotes the state obtained by pairing of the $l$th ancilla state to the committed qubit, $\sigma_\sb$ is the committed part of
the committed qubit-ancilla entangled pair. We have $(\sigma_\sb a_l)=\sigma_\sb\otimes a_l$ when the pairing is incorrect but is a properly qubit-ancilla
state when they match. From (\ref{eq:10}),
\begin{equation}
\label{eq:11}
n(\rho_0-\rho_1)=[(\sigma_0 a_{\bar{l}})-(\sigma_1 a_{\bar{l}})]\otimes_{l\neq \bar{l}} a_l+(\sigma_0-\sigma_1)\otimes_{l}a_l
\end{equation}

\noindent where $\bar{l}$ is the actual position of the committed qubit. Since $\sigma_0=\sigma_1=\frac{I}{2}$
for incorrect matching and $\|(\sigma_0-\sigma_1)\otimes a\|_1=\|\sigma_0-\sigma_1\|_1$ for all density operators $\sigma_0$, $\sigma_1$ and $a$, from (\ref{eq:11})
\begin{equation}
\label{eq:12}
\|\rho_0-\rho_1\|_1=\frac{1}{n}\|(\sigma_0 a_{\bar{l}})-(\sigma_1 a_{\bar{l}})\|_1=\frac{2}{n}
\end{equation}
which can be made arbitrarily small with large n. Equation (\ref{eq:12}) expresses exactly the intuitively obvious fact that Bob succeeds in cheating when and
only when he guesses correctly which original qubit the committed qubit is.
\\

\noindent Binding Proof for QBC1:

Since $\sigma_\sb=\frac{I}{2}$ without Bob's ancilla, the probability that Alice can determine the state of the committed qubit she chooses for commitment
is arbitrarily small, given by
$\frac{1}{M}$ for $M$ possible states on $C$ and is zero asymptotically. Without the possibility of entanglement cheating,
Alice can simply declare the bit she wants to open. In that situation $P^A_c$ is given by the inner product square of the two possible committed state.
With our choice $P^A_c=0$ since the two states for the two different $\sb$ values are orthogonal.
\\

Note that as in the discussion of QBC since the beginning, an $\epsilon$-concealing protocol can be made $\epsilon$-binding
in a sequence of committed qubits to obtain a single secure bit whenever $P^A_c$ is not too close to 1 for each original qubit. In the above QBC1,
such a sequence has also been indicated in our previous version in \cite{qbc3} for such purpose. It is not needed if the two
bit states are orthogonal or nearly orthogonal and if completely random qubit states on $C$ are supplied by Bob.

It remains to show that (\ref{eq:8}) can be checked and no security leak could occur during the checking process.
In contrast to the states sent in by Bob, it is more complicated to check (\ref{eq:8}) since Alice already committed by then, but it can be done as follow.
\\

\noindent Checking of Entanglement (\ref{eq:8}):

Alice would first send her ancillas of (8) to Bob with an entanglement basis unknown to him. Then Bob sends back Alice's
committed qubit to her who would turn it back to the original state by reversing her $U_b$. Then she sends back all qubits
to Bob who can thus (\ref{eq:8}).

We now show there can be no security compromise in the checking and each party must follow the prescription as all relevant states can be checked.
First, Bob can derive no information on which qubit he sent is committed without first knowing what qubit positions are indicated by what
ancilla state. The ancilla he so receives back from Alice is a totally random state to him.
Secondly, Alice must send in her entanglement ancilla and tell Bob later exactly what the total state is, as prescribed in the
checking. Third, that Bob then sends back the correct qubit can be checked by Alice via asking Bob to send back all the relevant states
in his possession which include his ancilla, Alice's committed qubit, and her ancilla that was sent him. Alice can then check similar to the beginning
check on Bob's $n$ qubit ancilla state. Finally, Alice must send back the proper states or else Bob cannot verify (\ref{eq:8}).

We have completed the security proof with proper operation procedure for QBC1. Assuming honest operation that we have shown can all be checked, the protocol can be simply summarized in the following:

\begin{center} \vskip 0.1in \framebox {
\begin{minipage}{0.9\columnwidth}
\vskip 0.1in \underline{PROTOCOL {\bf QBC1}}

{\small \begin{enumerate}
\item Bob sends Alice $n$-qubits, each randomly from a fixed great circle of the qubit Bloch sphere.
\item Alice forms (\ref{eq:8}) and modulates the first qubit by $U_{0}=R(\frac{\pi}{2})$ or $U_{1}=R(\frac{-\pi}{2})$ and sends it back to Bob.
\item Bob randomly chooses half of the qubits he sent and asks Alice to send them back for checking.
If Alice says it contains the committed one, Bob asks
to check the other half instead.
\item Alice opens by sending back all qubits and revealing everything; Bob
verifies.
\end{enumerate}
\vskip 0.1in }
\end{minipage}}
\end{center}
\vskip 0.11in

\section{Scope of QBC Possibility}\label{sec:scope}
It has long been known that a trusted third party or special relativistic effects can be used to establish secure bit commitment protocol
both classically and quantum mechanically. Furthermore, D'Ariano has suggested \cite{dariano09} that casuality or time order cannot be purified and
is built into quantum mechanics already in a way that would imply special relativity. If true this would imply quantum mechanics by itself would
ensure the possibility of secure QBC similar to Kent's relativistic protocol \cite{kent}. Cheung \cite{chau3} has recently proposed a
secure protocol on the basis of timing effect. In this paper, we show that quantum mechanics allows secure QBC without invoking causality
or timing, in a way that was first described in \cite{qbc3}.

The exact mechanism of how our QBC1 falls outside the standard impossibility proof is made clear in
section \ref{sec:newappr} above. There seem to be some vague claims of universal QBC impossibility in
ref \cite{dariano07} and \cite{chis}. Both papers are presented in unfamiliar mathematical
formulation of $C^\ast$-algebra or ``quantum comb'' with no translation into the usual formulation. In both of these new formulations, there is no clear
indication on exactly what would happen when one party is found cheating during protocol execution. Just aborting the
protocol is not enough as one party can keep on cheating as discussed in section \ref{sec:newappr}.
While the number of allowable protocol abortions may be bounded in \cite{chis}, cheating detection entails no penalty in any form.
More significantly, it appears there is no restriction put on the parties' entanglement purification and a private ancilla not to be checked is allowed, thus
excluding QBC1 in these formulations.

A most important point that is not addressed before in all the impossible proofs that claim universality
is \textit{what the proof is} that all possible QBC protocols have been included.
A general discussion of this issue can be found in \cite{yuen1}. A main point that has \textit{not} even been
made clear in \cite{yuen1} is that a `machine' formulation cannot capture all the possible protocols,
classical or quantum, that can be clearly formulated with ordinary natural language due to the `meaning' problem.
Specific intended meaning can be captured by a mechanical process, but not all possible
meaning in a general context. This is the situation of human knowledge that, I believe, would not be changed in the future.
In the present QBC issue, one manifestation of this situation is that there is no general
mathematical definition which captures all possible QBC protocols.

As a concluding remark, practical QBC protocols can be developed that can be proved secure within technological limits that are
unlikely to be removed in the foreseeable future. Entanglement across many qubits already by itself falls under these limits. Such implementable
protocols could be practically significant even if they are not unconditionally secure under the impractical assumption of ideal system devices and components.

\section*{Acknowledgments}

I would like to thank C.Y.~Cheung, G.M.~D'Ariano, and M.~Ozawa for very useful discussions.

\end{document}